\documentclass[twocolumn]{aastex631}

\usepackage{amsmath}
\usepackage{amssymb}
\usepackage{graphicx}

\shorttitle{Origin of Galactic Cosmic Rays}
\shortauthors{Author et al.}

\begin{document}

\title{A Minimal Interpretation of the Galactic Cosmic-Ray Proton and Helium Spectra from GeV to PeV Energies}

\author{Felix Aharonian}
\affiliation{Yerevan State University, 1 Alek Manukyan Street, Yerevan 0025, Armenia}
\affiliation{University of Science and Technology of China, 230026 Hefei, China}
\affiliation{TIANFU Cosmic Ray Research Center, 610000 Chengdu,  China}
\affiliation{Max-Planck-Institut for Nuclear Physics, P.O. Box 103980, 69029  Heidelberg, Germany}
\author{B. Theodore Zhang}
\affiliation{Institute of High Energy Physics, Chinese Academy of Sciences, 100049 Beijing, China}
\affiliation{TIANFU Cosmic Ray Research Center, 610000 Chengdu, China}

\begin{abstract}

High-precision measurements of the cosmic-ray (CR) proton and helium spectra have revealed significant deviations from a simple power law, characterized by multiple spectral features, including a hardening above $\sim$100~GeV, a broad excess in the multi-TeV range, and a pronounced structure at PeV energies. 
We propose a minimal phenomenological two-component cosmic-ray framework that consistently reproduces the main observed features of the proton and helium spectra across six decades in energy, with agreement at the level of $\sim 10 \%$  or better over most of the explored energy range.
In this scenario, the spectral complexity arises from a transition between two Galactic CR populations in the 10~TeV–1~PeV energy range. The low-energy proton population exhibits a sharp cutoff at tens of TeV, while a second, higher-energy population emerges and dominates above 100~TeV, terminating with a smooth exponential cutoff at $\sim$6.5~PeV.
The same two-component model applied to CR helium, with a slightly harder first component extending effectively to several hundred TeV and a second component that scales with the proton spectrum in magnetic rigidity, provides a consistent description of both the helium spectrum and the  p/He ratio.
This framework reproduces the main observed spectral features of CR protons and helium without invoking contributions from nearby sources or non-standard assumptions about CR acceleration or propagation. Recent gamma-ray observations of supernova remnants, star-forming regions, and microquasars offer plausible astrophysical sites for these two CR components.

\end{abstract}


\section{Introduction}
Galactic cosmic rays (CRs) were long viewed as a uniform population characterized by 
a featureless power-law spectrum ending  at energies 
of a few PeV. The concept of the origin of Galactic CRs evolved from the 
initial SN hypothesis of \cite{BaadeZwicky1934} to the modern SNR paradigm \citep[e.g.,][] {GinzburgSyrovatskii1964}, which 
attributes the bulk of the Galactic CR flux to particles accelerated in supernova remnants (SNRs) via
the so-called process of diffusive shock acceleration (DSA; see \citealt{MalkovDrury2001} for a review).

In its simplest test-particle formulation, DSA at strong shocks predicts a
power-law source spectrum with index $\Gamma = 2$.
More realistic, time-dependent treatments, including  for shock obliquity
deceleration, evolving magnetic turbulence, and energy-dependent particle escape, 
predict steeper source spectra, $\Gamma \simeq 2.3$--2.4. Subsequent
 energy-dependent propagation of CRs in Galactic magnetic fields further steepens the
CR energy distribution, resulting in an observed spectrum close to $E^{-2.7}$
\citep[e.g.,][]{Strong2007Review,Grenier2015Review,Gabici2019Review}.

Within this framework, the spectral feature known as `knee', first identified in the late 1950s
\citep{Kulikov1959}, is widely interpreted as marking the termination of the Galactic CR
component at the maximum energy achievable in young SNRs. However, conservative
theoretical treatments show that, under standard conditions, proton energies
are unlikely to exceed $100~\mathrm{TeV}$ \citep{LagageCesarsky1983a}. 
This tension has motivated extensions of the standard
DSA  concept by proposals  such as magnetic-field amplification 
driven by nonlinear CR-induced instabilities \citep{Bell2004}, 
or by rigidity-dependent interpretations in which
the `knee' reflects the cutoff of heavy  nuclei, whereas protons terminate at lower
energies \citep{Horandel2003}.

Recent high-precision measurements of the Galactic CR proton spectrum have
significantly refined this picture. Data from space- and ground-based
experiments, spanning nearly six decades in energy and largely free from
nuclear-composition ambiguities reveal clear deviations from a simple
power-law behaviour. The spectrum exhibits a gradual hardening beginning at
$\sim 200~\mathrm{GeV}$, followed by saturation and subsequent softening in the
multi-TeV range (see Fig.~\ref{fig1}), resulting in a relatively narrow
spectral feature at a few  tens of TeV, hereafter referred to as the
`multi-TeV bump'.

Although the energy range between tens and a few hundred TeV is affected
by non-negligible statistical and systematic uncertainties, 
current data
indicate a distinct spectral turnover near $100~\mathrm{TeV}$. At higher
energies, measurements extending into the PeV domain reveal an additional broad
spectral structure characterized by a pronounced maximum at several PeV,
hereafter referred to as the `PeV bump'. Identification of the proton component extending to at least 
10~PeV resolves the degeneracy associated with heavy nuclei
contribution and suggests the existence of CR proton PeVatrons.
The presence of multiple distinct spectral structures cannot be adequately addressed within the standard Galactic CR framework and has prompted alternative interpretations,
including specific acceleration scenarios, revised propagation models, and substantial local effects.  

The spectral hardening can be explained by the modification of the shock precursor by CR pressure, resulting in a concave CR energy spectrum \citep{MalkovDrury2001}. Another possibility related to particle acceleration in SNRs is the enhanced contribution of a polar-cap CR component,  produced by supernova explosions within the winds of massive progenitor stars \citep{Biermann2010}. 
Alternatively, the spectral hardening could be linked to the specifics of CR propagation outside the accelerators \citep[e.g.,][]{Tomassetti2012, Blasi_hardening2012, Genolini2017, Sarah_hardening}. It has been shown, for example, that this effect can be explained by assuming a spatial change in CR transport properties, specifically the diffusion coefficient, across different regions of the interstellar medium  \citep{Tomassetti2012}. 

While both acceleration and propagation scenarios remain viable, it has been argued that measurements of cosmic-ray secondaries favor a diffusive propagation origin for the spectral hardening over acceleration-based explanations \citep{Genolini2017}. However, recent DAMPE measurements of the B/C ratio \citep{DAMPE_BtoC} suggest that fundamental modifications to the propagation models themselves may be required, rather than mere parameter adjustments; for example, through nested diffusion, whereby cosmic rays spend a non-negligible fraction of their lifetime in the vicinity of their sources \citep{Yang_FA2025}.

The third approach to interpreting the “multi-TeV bump” assumes that it is produced by one or more nearby sources whose contributions peak at tens of TeV \citep{Thoudam2012, Liu:local, DAMPE:bump, Yuan:2025xqt}. The hypothesis of a dominant local source, such as a middle-aged supernova remnant (SNR), is indeed a viable scenario \citep{AhAtVl1995, ErlWolf1997}. Moreover, when modeling the total CR flux from Galactic sources, it is important to distinguish between nearby sources (typically within a few hundred parsecs) and the integrated contribution of a continuous distribution of distant sources. Depending on its power, age, distance, and the diffusion coefficient, a local source may contribute significantly, imprinting a distinct feature on the otherwise smooth diffuse spectrum \citep{AhAtVl1995}.

Another interesting and distinct version within this `local effect'  category suggests that the `multi-TeV' bump results from spectral distortions induced by the re-acceleration of the Galactic CR background by a weak shock associated with a nearby star,  located only a few parsecs from the Sun \citep{MalkovMoskalenko2021}.

While the above models can account, with more or less success, for the spectral features revealed in recent CR observations, we propose a simple phenomenological framework in which the superposition of two broadband Galactic CR components - characterized by different power-law indices and cutoff energies separated by two orders of magnitude  - naturally reproduces the observed spectral features of protons, helium, and, presumably, heavier species. This approach requires neither nonstandard acceleration nor propagation scenarios nor local effects such as the presence of a dominant nearby source. In this paper, we restrict the analysis to proton and helium species.

Finally, we emphasize that the proposed phenomenological framework is intended to describe the global structure of the CR spectra rather than to provide a precision-fit model of every spectral detail. Nevertheless, over more than six decades in energy, the model reproduces the observed CR proton and helium spectra with agreement typically at the level 
of $\sim 10 \%$ or better.

\section{Observations} 

\begin{figure}[htbp]
    \centering
    \includegraphics[width=\linewidth]{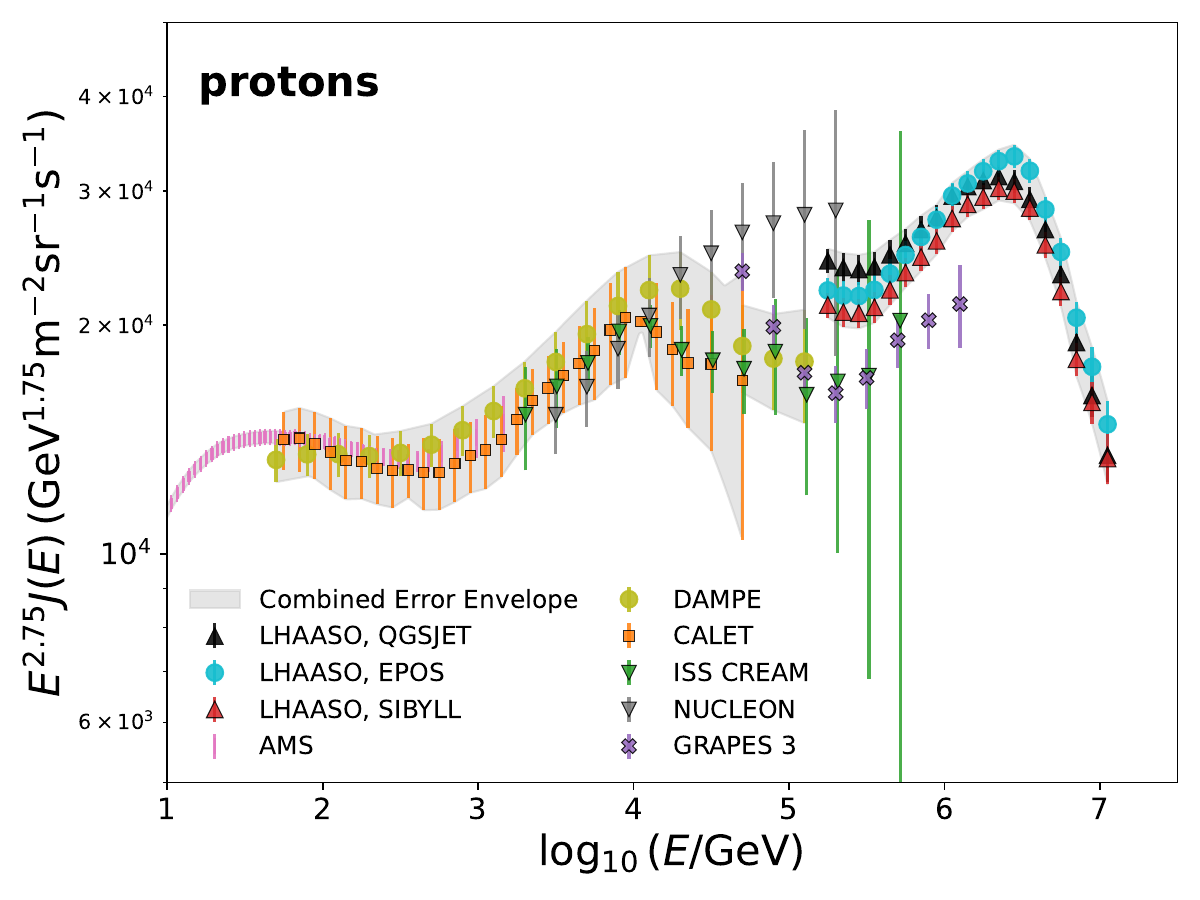}
    \includegraphics[width=\linewidth]{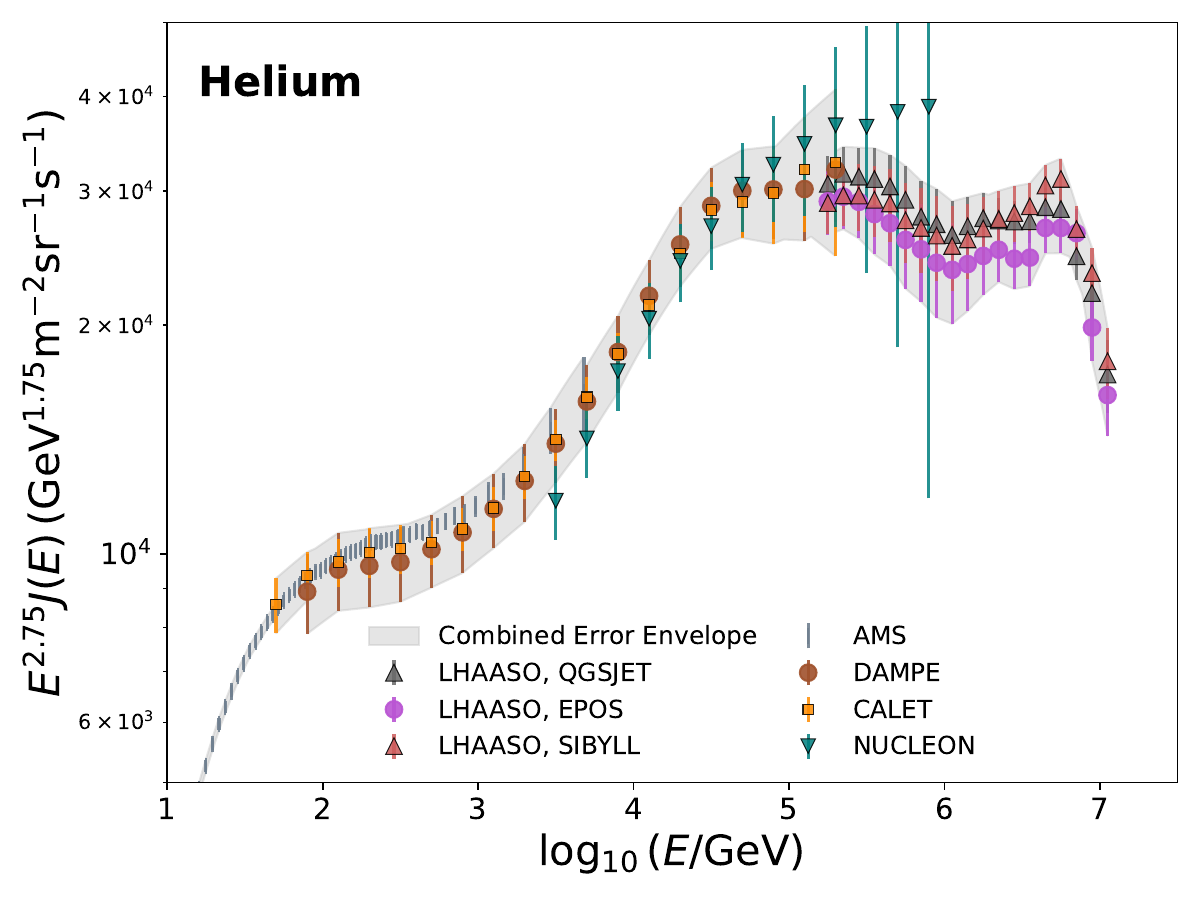}
    \caption{Cosmic-ray proton (a) and helium (b) energy spectra, multiplied by $E^{2.75}$, as a function of primary energy $\log_{10}(E/\mathrm{GeV})$.
\textbf{(a)} Proton measurements compiled from AMS~\citep{Aguilar2015Proton}, CALET~\citep{CALET_2022_proton}, DAMPE~\citep{DAMPE:2025opn}, ISS-CREAM~\citep{Choi:2022aht}, GRAPES-3~\citep{GRAPES2024}, NUCLEON~\citep{Gorbunov:2018stf}, and LHAASO~\citep{LHAASO:2025byy}.
\textbf{(b)} Helium measurements compiled from AMS~\citep{AMS:2017seo}, CALET~\citep{CALET:2023nif}, DAMPE~\citep{DAMPE:2025opn}, NUCLEON~\citep{Gorbunov:2018stf}, and LHAASO~\citep{LHAASO:2025mlf}.
For the LHAASO data in both panels, fluxes reconstructed using three high-energy hadronic interaction models ({\sc QGSJET-II-04}, {\sc EPOS-LHC}, and {\sc SIBYLL~2.3d}) are shown separately.
The shaded gray band represents a combined uncertainty envelope defined 
by the spread of the AMS-02, DAMPE, CALET, and LHAASO measurements. It 
illustrates the overall experimental dispersion rather than a formal 
statistical uncertainty.
\label{fig:proton_helium_spectrum}}
\end{figure}


The unambiguous separation of distinct CR species (protons, helium, and heavier nuclei) is a fundamental requirement for understanding the origin and propagation of Galactic cosmic rays. Precise measurements of protons and helium are especially critical, as they constitute the most abundant primary species and provide the highest-statistics data for testing acceleration and propagation models.

After decades of steady progress in CR measurements, major advances were achieved in the 2010s with results from PAMELA in the GeV range and from CREAM at TeV energies.
These observations revealed previously unrecognized spectral features, in particular showing that the fluxes of CR species cannot be described by a single power law.

Subsequently, AMS-02 confirmed the spectral hardening with unprecedented precision and extended the measurements up to nearly 2~TV in rigidity. Later, several space- and balloon-borne detectors, in particular NUCLEON~\citep{Gorbunov:2018stf}, 
ISS-CREAM~\citep{Choi:2022aht}, CALET~\citep{CALET_2022_proton}, and DAMPE~\citep{DAMPE:2025opn}, performed independent measurements in the multi-TeV range, indicating the presence of a broad spectral bump between $\sim$10 and $100$~TeV.

Over the last decade, significant progress has also been achieved at PeV energies by ground-based detectors, including KASCADE-Grande~\citep{KASKADE_spe}, GRAPES-3~\citep{GRAPES2024}, IceTop~\citep{IceTop}, and LHAASO~\citep{LHAASO:2025mlf}. Using a multi-component air-shower approach, the LHAASO collaboration has extracted the CR proton population with unprecedented purity, significantly reducing the systematic uncertainties associated with the hadronic interaction models used for primary mass identification and energy reconstruction.
Measurements of comparable quality have also been obtained for helium, the second-most-abundant CR species. 

Figure~\ref{fig:proton_helium_spectrum} presents a compilation of CR proton and helium spectra, multiplied by $E^{2.75}$, over a wide energy range from 10~GeV to 10~PeV. At low,
sub-TeV energies, the proton and helium spectra are measured with extremely high precision by AMS-02  and exhibit a gradual spectral hardening with energy. This feature is consistently confirmed by CALET and DAMPE,  which extend direct measurements into the TeV range.  At low energies,  below 200 GeV, one can see a spectral change --  a steepening both in the proton and helium spectra. The trend of steepening is especially 
apparent in the proton spectrum, which, between 50 and 200 GeV, exhibits a concave shape. Below 50 GeV, the measured fluxes are affected by solar modulation, and recovering the initial spectrum requires dedicated treatment and modeling.

In the multi-TeV region, both proton and helium spectra exhibit a broad excess --- a ``multi-TeV bump'' --- relative to a single power-law behavior, although its amplitude and detailed shape vary among experiments. Measurement uncertainties evolve significantly with energy and measurement technique. In the GeV--TeV range, direct measurements achieve percent-level precision, with systematic differences between experiments becoming apparent when spectral features are emphasized.

Above $\sim$10~TeV, and especially above 100~TeV, direct measurements are increasingly affected by limited statistics, leading to larger uncertainties for individual experiments and greater dispersion among datasets. This is particularly evident in the proton flux measurements reported by NUCLEON~\citep{Gorbunov:2018stf}. The dispersion is less pronounced in the helium measurements.

In Fig.~\ref{fig:proton_helium_spectrum}, we show the envelopes of the combined uncertainties in the proton and helium fluxes as shaded gray bands derived from the AMS-02, DAMPE, CALET, and LHAASO data, which are used in this work as representative datasets in the GeV, TeV, and PeV energy ranges. These bands illustrate the relatively small and well-controlled experimental uncertainties over the entire energy range from GeV to PeV.

Around 100~TeV, the uncertainty bands broaden significantly, reflecting the effects of limited statistics in direct measurements. Above this energy, the ground-based measurements of the proton and helium fluxes by LHAASO provide the most statistically significant proton and helium measurements currently available.
Furthermore, the hybrid detector configuration of LHAASO enables a particularly robust separation of proton and helium nuclei, reducing the systematic uncertainties associated with composition-dependent air-shower reconstruction.
In this regime, uncertainties associated with the modeling of hadronic interactions become increasingly important. Even so, the dispersion among spectra reconstructed with different hadronic interaction models remains modest and is comparable to the uncertainties in direct measurements at GeV and TeV energies.
Consequently, the reconstruction of the proton and helium spectra remains remarkably stable above $\sim$1~PeV, with variations between different hadronic interaction models limited to $\sim 10 \%$ or less (Fig.~\ref{fig:proton_helium_spectrum}).

In summary, the wealth of high-precision measurements accumulated over the past decade has established a complex structure in the CR proton and helium spectra. These structures include a spectral hardening at several hundred GeV, a broad excess in the multi-TeV region, and a softening toward PeV energies associated with the knee. While the amplitude and detailed shape of these features vary among experiments, and the associated uncertainties increase with energy -- particularly due to limited statistics and model-dependent effects in air-shower reconstructions -- their presence is consistently supported by multiple independent measurements employing diverse detection techniques. This overall consistency strongly suggests that the observed spectral features are of physical origin rather than artifacts of individual instruments or analysis methods.  The analysis presented below focuses on a representative subset of proton and helium measurements reported by the AMS-02, DAMPE, LHAASO, and IceTop collaborations. 
These experiments were selected for their complementary energy coverage, combining percent-level precision at GeV--TeV energies with robust measurements in the multi-TeV to PeV regime, as well as for their use of independent detection techniques that provide important cross-checks of the observed spectral features.


\section{The framework}

We describe the energy spectra of Galactic CR species as the superposition of two distinct components, each characterized by a power-law distribution with a high-energy cutoff. The two components are introduced phenomenologically, without specifying their physical origin, in order to provide a minimal description of the observed spectra across the entire GeV--PeV energy range. In this work, we focus on the proton and helium components.

We describe the energy spectra of Galactic CR species as the superposition of two distinct components, each characterized by a power-law distribution with a high-energy cutoff. These components are introduced phenomenologically, without specifying their physical origin, to provide a minimal description of the observed spectra across the entire GeV--PeV energy range. In this work, we focus specifically on the proton and helium populations.

We write the total proton flux as
\begin{equation}
J_{\rm tot}(E) = J_1(E) + J_2(E),
\end{equation}
where $J_1(E)$ denotes the low- to intermediate-energy component dominating at
GeV and TeV energies,  and $J_2(E)$ represents the second, harder component 
that becomes significant at multi-TeV energies and extends to much higher energies.

Both components are parameterized using the same functional form,
\begin{equation}
J_i(E) = A_i \, E^{-\Gamma_i}
\exp\!\left[-\left(\frac{E}{E_{i,0}}\right)^{\beta_i}\right],
\label{eq:Ji}
\end{equation}
where $A_i$ is the normalization, $\Gamma_i$ the power-law spectral index, $E_{i,0}$ the
characteristic cutoff energy, and the parameter $\beta_i$ controls the sharpness of the cutoff 
for $i=1$ and $i=2$ components. The adopted functional form is not intended
to offer a  description of acceleration or transport processes, 
but rather to extract the characteristic energy
scales, spectral slopes, and cutoff energies that are directly implied by the
observations. It is sufficiently general to provide a flexible phenomenological
description applicable to a broad range of astrophysical scenarios.

Typically, the superposition of two broad spectral components—such as two power-law functions with different indices—results in a smooth, gradual transition in the total spectrum where their contributions are comparable. However, the measured CR proton spectrum deviates from this expectation. Instead, the data reveal a relatively weak but visible excess in the multi-TeV range, followed by a pronounced and sharp turnover within a narrow energy interval around 100 TeV. Similar, though somewhat less prominent, features are observed in the CR helium spectrum (see Fig. \ref{fig1}).  

To accurately reproduce the observed turnover within such a narrow energy interval, the low-energy component must be suppressed above its characteristic cutoff energy. In practice, this necessitates a cutoff significantly steeper than a simple exponential, for example, a super-exponential form with $\beta_1 > 1$ in Eq.~(\ref{eq:Ji}).
Consequently, the contribution of the first component becomes strongly suppressed at energies exceeding $100~\mathrm{TeV}$. 

At higher energies, the observed spectrum is dominated by the second component, allowing its spectral shape to be constrained directly by measurements in the PeV range, where the first component is negligible. Once this high-energy term is determined, it can be extrapolated to lower energies. Subtracting it from the total measured CR flux then isolates the spectrum of the low-energy component.

\section{Results}

\subsection{CR  proton spectrum}

\begin{figure}[t]
  \centering
   \includegraphics[width=\linewidth]{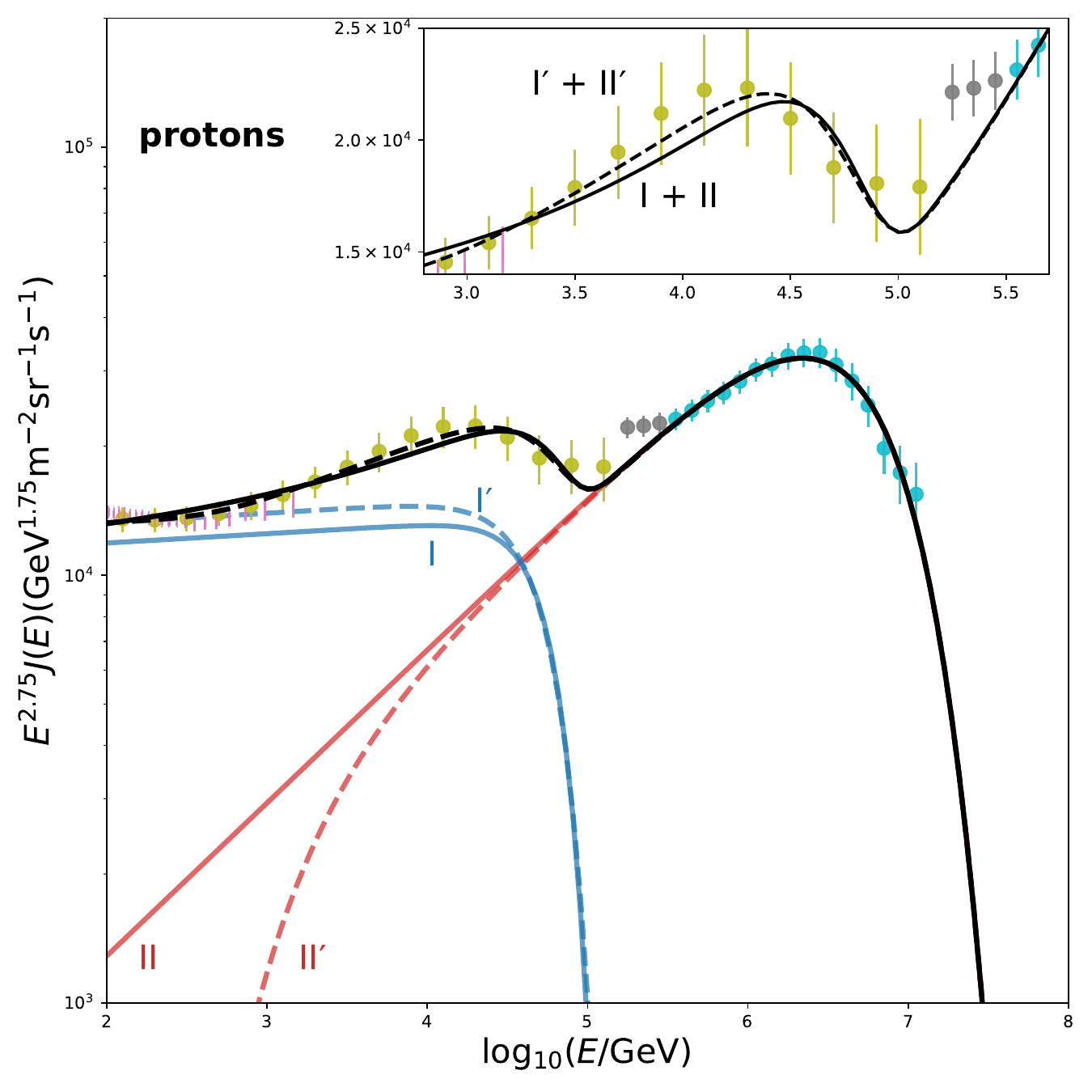}
  \caption{\label{fig1}
CR proton spectrum multiplied by $E^{2.75}$, with fit curves compared to 
measurements  from AMS-02\citep{Aguilar2015Proton}, DAMPE \citep{DAMPE:2025opn}, and LHAASO \citep{LHAASO:2025byy}.  The blue and red curves represent the contributions of the first (I) and second (II) CR components, respectively. Two alternative assumptions for the extrapolation of the second CR component toward low energies are shown: (a) a pure power-law extrapolation (curves $\rm II$)
and (b) a scenario with low-energy hardening of the second component (curves $\rm II^{\prime}$),  as discussed in the text. 
The solid and dashed black curves show the total proton spectrum 
in the two-component CR framework for the (a) and (b) scenarios, respectively.  
The corresponding best-fit parameters are listed in Table~\ref{tab:spectral_fits_1}. The inset highlights the spectral structure in the multi-TeV energy range. 
  }
\end{figure}

Using the two-component framework described above, we fit the measured proton spectrum from $100~\mathrm{GeV}$ to $10~\mathrm{PeV}$. At the highest energies, above $\sim 300~\mathrm{TeV}$, the observed spectrum is dominated by the second component, allowing its spectral parameters to be constrained directly by measurements from LHAASO Collaboration. In this energy range, the high-energy contribution is well described by a power-law spectrum with index $\Gamma_2 \simeq 2.4$ and an exponential cutoff ($\beta_2 = 1$) at $E_{2,0} \simeq 6.2~\mathrm{PeV}$ (see Table~1).\footnote{To avoid implying unrealistically high accuracy, the parameter values quoted in the text are rounded, whereas Table~1 lists the formal best-fit values obtained from the fitting procedure.}


\begin{deluxetable*}{lccccc}
\tablecaption{Fit parameters for the two-component CR proton
model\label{tab:spectral_fits_1}}
\tablehead{
     \colhead{Model} &
     \colhead{$A$} &
     \colhead{$\Gamma$} &
     \colhead{$E_{0}$} &
     \colhead{$\beta$} &
     \colhead{$E_{\rm min}$} \\
     \colhead{} &
     \colhead{($\rm GeV^{1.75}~m^{-2}~s^{-1}~sr^{-1}$)} &
     \colhead{} &
     \colhead{(TeV)} &
     \colhead{} &
     \colhead{(TeV)}
}
\startdata
\multicolumn{6}{l}{First component} \\
${\rm I}$   & $14500_{-1900}^{+2500}$ & $2.73 \pm 0.02$ &
$68_{-14}^{+18}$ & $2.55_{-0.90}^{+1.32}$ & \nodata \\
${\rm I}'$  & $16200_{-2700}^{+4200}$ & $2.73_{-0.03}^{+0.02}$ &
$63_{-12}^{+16}$ & $2.18_{-0.74}^{+1.31}$ & \nodata \\
\hline
\multicolumn{6}{l}{Second component} \\
${\rm II}$  & $34700 \pm 1600$ & $2.39 \pm 0.05$ & $6200_{-800}^{+900}$
& 1 (fixed) & \nodata \\
${\rm II}'$ & $34700$ (fixed) & $2.39$ (fixed) & $6200$ (fixed) &
1 (fixed) & $0.89_{-0.57}^{+1.30}$ \\
\enddata
\tablenotetext{}{
The fitting function $E^{2.75} J(E)$ is presented in the form
\[
E^{2.75} J(E) = A \left(\frac{E}{E^\star}\right)^{2.75-\Gamma}
\exp\!\left[-\left(\frac{E}{E_{0}}\right)^{\beta}\right]
\exp\!\left[\left(-\frac{E_{\rm min}}{E}\right)\right],
\]
where $E_{0}$, $\Gamma$, and $\beta$ are the fit parameters listed in
the table,
$E^\star = 10^{3}~\mathrm{TeV}$ is a fixed pivot energy. 
The normalization parameter $A$ is given in units of
$\mathrm{GeV^{1.75}\,m^{-2}\,s^{-1}\,sr^{-1}}$, corresponding
to the
representation shown in Fig.\ref{fig1}.
The first CR component is obtained from a joint fit to the DAMPE and
LHAASO data,
while the second component is constrained exclusively using LHAASO data above
$E > 300~\mathrm{TeV}$.
}
\end{deluxetable*}


Extrapolating this component to lower energies and subtracting it from the measured data isolates the low-energy proton component. This component is characterized by a spectral index $\Gamma_1 \simeq 2.7$, a cutoff energy $E_{1,0} \simeq 67.6~\mathrm{TeV}$, and a super-exponential cutoff parameterized by $\beta_1 \simeq 2.5$ (see Table~1). The resulting superposition of the two components reproduces the observed  multi-TeV bump, the sharp turnover near $100~\mathrm{TeV}$, and the broad spectral structure in the PeV range.

The fit parameters presented in Table~\ref{tab:spectral_fits_1} were derived from a joint analysis of the DAMPE and LHAASO data sets. AMS-02 data were excluded from the fitting procedure due to their excellent agreement with DAMPE measurements in the overlapping energy range, from approximately $40~\mathrm{GeV}$ to $1.8~\mathrm{TeV}$. The power-law index of the derived first component, $\gamma \approx 2.74$, is lower by $\approx 0.1$ than that of the AMS-02 spectrum around $200~\mathrm{GeV}$. While the measured spectrum exhibits gradual hardening, this component remains approximately constant up to tens of TeV. Consequently, when modeling CR acceleration and transport in Galactic magnetic fields, the derived first proton component should be used rather than the measured spectrum, irrespective of its higher measurement precision.

A modest deviation is observed in the LHAASO data below $\sim 300~\mathrm{TeV}$. These points indicate an unusually flat local spectral behavior that is difficult to reconcile with current models or general expectations from diffusive acceleration and propagation scenarios. While this could, in principle, suggest a narrowly confined spectral component, the deviation in absolute flux does not exceed 
$10 \% $. Furthermore, this energy interval lies close to the LHAASO detection threshold, where systematic uncertainties in energy reconstruction and detector response are most significant. Consequently, we do not attempt an independent fit to these data, but instead focus on the global spectral behavior robustly described by the two-component framework. Should future measurements confirm a feature confined to the $100$--$300~\mathrm{TeV}$ range, its interpretation within standard models would present a significant theoretical challenge.

The extrapolation of the high-energy component to GeV energies is not uniquely constrained by the available data. While a simple power-law extrapolation is consistent with standard acceleration or propagation models, alternative low-energy behaviors remain plausible. For illustration, we consider a variant in which the second component exhibits low-energy suppression, as shown by the curve labeled $\mathrm{II'}$ in Fig.~\ref{fig1}. Specifically, we assume that below the TeV scale, the spectrum deviates from a pure power law due to an additional suppression term of the form $\exp[-E_{\rm min}/E]$. 

In this scenario, a slight adjustment to the normalization of the first proton component is sufficient to accurately describe the data, resulting in only minor changes to the fitted parameters (see Table~\ref{tab:spectral_fits_1}). This demonstrates the robustness of the two-component interpretation with respect to variations in the low-energy behavior of the second proton component. In contrast, a sharper low-energy cutoff of the second component would be incompatible with a power-law first component of fixed, energy-independent index, as it would fail to reproduce the gradual hardening observed above $200~\mathrm{GeV}$.

The parameter values listed in Table~\ref{tab:spectral_fits_1} should be regarded as representative examples rather than unique solutions. They illustrate the minimal constraints imposed by the observational data. Nevertheless, significant deviations from these values would degrade the fit quality. In particular, both the power-law index and the sharp cutoff of the first component below $\sim 100~\mathrm{TeV}$ are tightly constrained by observations. The parameters of the second component are well constrained at PeV energies, with the main remaining uncertainty arising from its extrapolation to lower energies.

\begin{figure}[t]
  \centering
  \includegraphics[width=\linewidth]{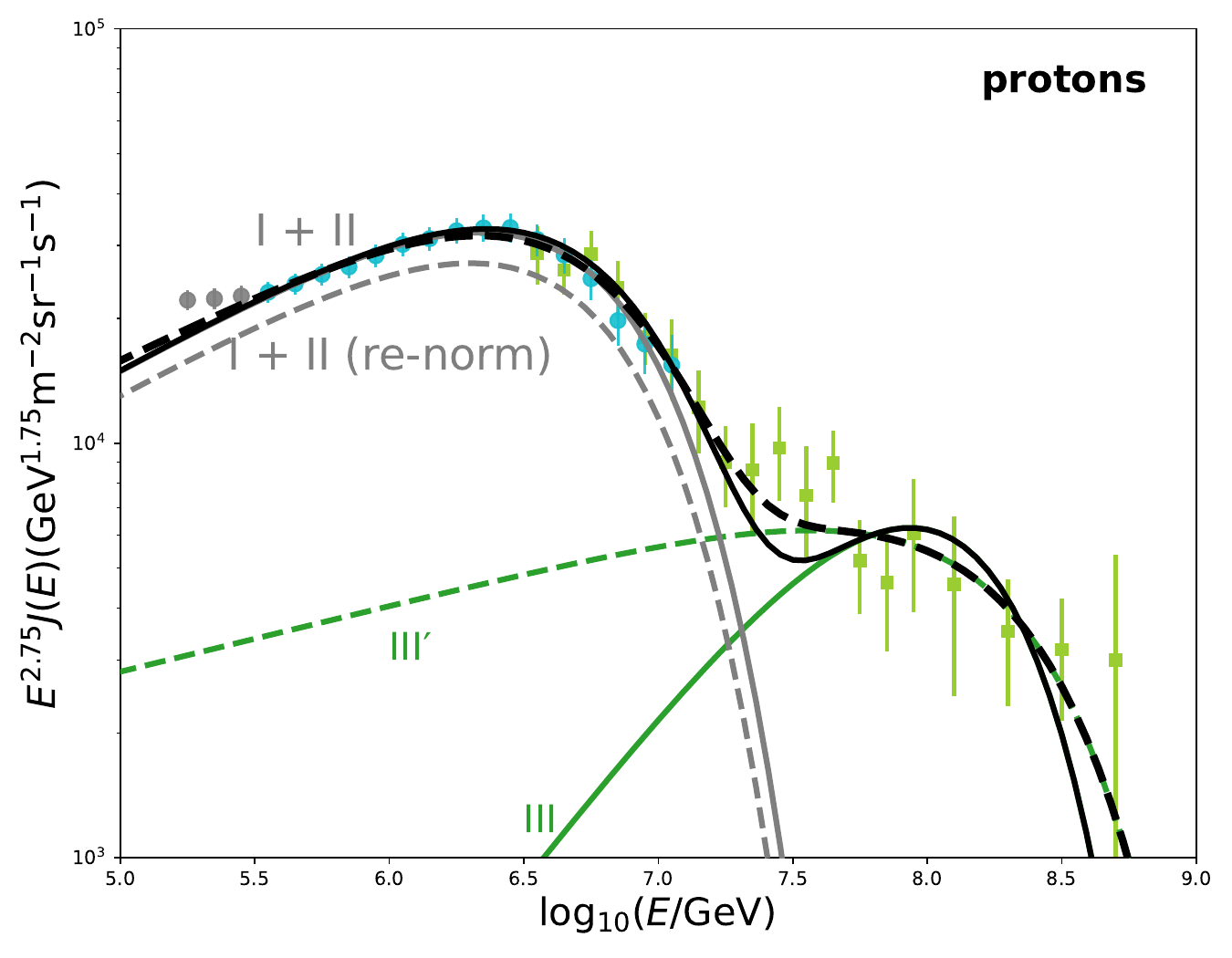}
  \caption{\label{fig2}
  CR proton spectrum multiplied by $E^{2.75}$, with fit curves compared to 
 the  LHAASO \citep{LHAASO:2025byy} and IceTop\citep{IceTop} measurements. 
The grey and green curves show the contributions of the first and second ($\rm I + II$) and third ($\rm III$) proton components, respectively.
 The sum of the first, second, and third components is shown by the black curves.
 The solid curves correspond to the case when the 
 flux of $\rm I + II$ is fixed and coincides with the flux in Fig.\ref{fig1} for the scenario (a). 
 The dashed curves are obtained from the joint LHAASO and IceTop data fit. 
 The corresponding best-fit parameters are listed in Table~\ref{tab:spectral_fits_2}.   }
\end{figure}


\begin{deluxetable}{lccc}
\tablecaption{Fit parameters for the third CR proton component \label{tab:spectral_fits_2}}
\tablehead{
    \colhead{Model} &
    \colhead{$A$} &
    \colhead{$\Gamma_3$} &
    \colhead{$E_{\rm max}$} \\
    \colhead{} &
    \colhead{($\rm GeV^{1.75}\,m^{-2}\,s^{-1}\,sr^{-1}$)} &
    \colhead{} &
    \colhead{PeV}
}
\startdata
\hline
\multicolumn{4}{l}{Third component\tablenotemark{a}} \\
Fixed $\rm II$  & $340_{-170}^{+280}$ & $1.91_{-0.21}^{+0.19}$ & $102_{-26}^{+43}$ \\
\hline
\multicolumn{4}{l}{Third component\tablenotemark{b}} \\
Free $\rm II$ & $4100 \pm 2000$ & $2.59_{-0.19}^{+0.12}$ & $229_{-85}^{+178}$ \\
\enddata
\tablenotetext{a}{Third component is fitted assuming the second component is fixed to its best-fit 
parameters ($\rm II$).}
\tablenotetext{b}{Third component is obtained from a joint fit in which the normalisation and cutoff energy of the second component are allowed to vary, while its spectral index is fixed at $\Gamma_2=2.4$. This yields $E_{2,0} = 5.6_{-0.4}^{+0.3}$~PeV and $A_2 = (3.02 \pm 0.28)\times10^4\rm~ GeV^{1.75}\,m^{-2}\,s^{-1}\,sr^{-1}$ for the second component.}
\end{deluxetable}


\begin{figure*}[htbp]
    \centering
    \includegraphics[width=0.75\textwidth]{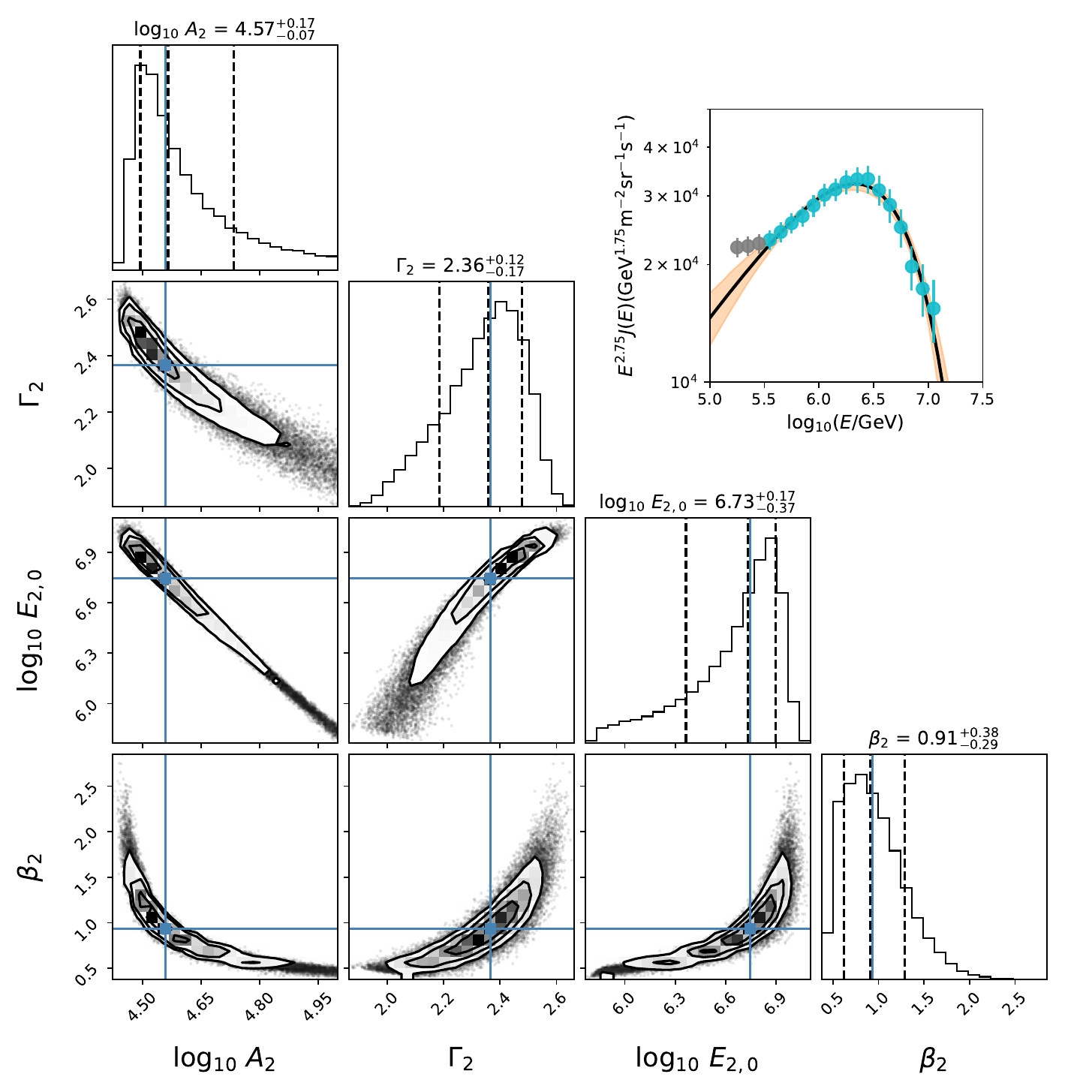}
    \caption{\textbf{Main panel:} Corner plot showing the posterior probability distributions for the model parameters for the second component. The sharpness parameter $\beta_2$ is treated as a free index within the exponential cutoff to characterize the spectral curvature. \textbf{Inset:} The observed spectrum compared with the best-fit model (solid black line). The shaded orange region represents the $1\sigma$ (16th--84th percentile) uncertainty envelope derived from 200 MCMC posterior samples, demonstrating the model's consistency with the data.}
    \label{fig3}
\end{figure*}

The detailed shape of the spectrum in the multi-PeV range may be influenced by an additional contribution from a third CR proton component at energies of several tens to a few hundred PeV. As shown in Fig.~\ref{fig2}, the fluxes reported by the IceTop Collaboration \citep{IceTop} in this energy range lie significantly above the extrapolation of the two-component spectrum, suggesting the emergence of a new CR proton component. Its origin may be Galactic or extragalactic, but this distinction is not essential for the purposes of this paper.

To investigate the interplay between the second and third proton components, we performed a joint fit to the combined LHAASO and IceTop data from $0.3$ to $\simeq 500~\mathrm{PeV}$. As a first step, we assume that the contribution of the third component below $10~\mathrm{PeV}$ is negligible, so that the combined spectrum of the first and second components remains unchanged. In practice, this implies fixing the second component, as the contribution of the first component in this energy range is negligible. Specifically, in the fitting procedure, we fix the second component to $\mathrm{II}$, corresponding to scenario (a) in Fig.~\ref{fig1}. Under this assumption, the resulting third component is characterized by a very hard power-law spectrum with $\Gamma_3 \simeq 1.9$ and an exponential cutoff at $E_{0,3} \simeq 100~\mathrm{PeV}$ (see Fig.~\ref{fig2} and Table~\ref{tab:spectral_fits_2}).

The agreement of this model with the data in the transition region around $\sim 30~\mathrm{PeV}$ is only marginal. The fit improves when the parameters of the second component are allowed to vary together with those of the third component. Although measurements in this energy interval are affected by large statistical uncertainties, they favor a softer proton spectrum with $\Gamma_3 \gtrsim 2.6$ and a significantly higher cutoff energy.

In this joint-fit procedure, the parameters of the second component change only modestly: its normalization is slightly reduced, and its cutoff energy decreases to $E_{2,0} \simeq 5.8~\mathrm{PeV}$, compared with $\simeq 6.3~\mathrm{PeV}$ in the pure two-component fit (see Fig.~\ref{fig2} and Table~\ref{tab:spectral_fits_2}). This stability can be readily understood. In the representation $E^{2.75} J(E)$, the flux at tens of PeV is substantially lower than that at a few PeV, implying that the third component cannot strongly affect the derived spectrum in the PeV range. While in principle the contribution of the third component below $\sim 10~\mathrm{PeV}$ could be increased by assuming an even steeper spectrum, such a scenario would contradict the IceTop measurements at the highest energies. We therefore conclude that the spectrum of the second component, as inferred from the LHAASO data, is stable and largely unaffected by the presence of the third CR component.

In summary, the Galactic CR proton spectrum from GeV to PeV energies is well described by the superposition of two broad components with power-law indices $\Gamma_1 \simeq 2.7$ and $\Gamma_2 \simeq 2.4$, extending up to $\sim 65~\mathrm{TeV}$ and $\sim 6.5~\mathrm{PeV}$, respectively. The first component exhibits a sharp termination, requiring a super-exponential cutoff with $\beta_1 \geq 2$, whereas the second is well described by a simple exponential cutoff. Their overlap near $\sim 100~\mathrm{TeV}$ naturally produces the observed multi-TeV bump, followed by a sharp spectral turnover, in good agreement with the data. Within this framework, the gradual hardening of the spectrum from hundreds of GeV to tens of TeV also arises naturally from the superposition of the two components, without invoking non-standard acceleration or propagation effects.

In general, the two-component approach can be naturally extended to other CR species. However, the correspondence between protons and heavier nuclei is not straightforward, as they may be predominantly produced in different types of cosmic accelerators. Moreover, even if protons and nuclei originate from the same class of sources, the superposition of the two components is not expected to be identical for different species, except in the unlikely case where both components share identical chemical composition and spectral shape at a given rigidity.

We next consider the CR helium spectrum, which provides a complementary test of the two-component scenario, as differences in composition and spectral behavior between protons and helium can constrain the relative contributions of the two components. Since helium has a different mass-to-charge ratio ($A/Z = 2$) than protons, any spectral differences at the same rigidity would point to either different source environments (e.g., hydrogen-rich vs.\ helium-rich SNRs) or distinct injection efficiencies at the shock front.

\begin{deluxetable*}{lccccc}
\tablecaption{Fit parameters for the two-component CR helium model\label{tab:spectral_fits_he}}
\tablehead{
     \colhead{Model} &
     \colhead{$A$} &
     \colhead{$\Gamma$} &
     \colhead{$E_{0}$} &
     \colhead{$\beta$} &
     \colhead{$E_{\rm min}$} \\
     \colhead{} &
     \colhead{($\rm GeV^{1.75}~m^{-2}~s^{-1}~sr^{-1}$)} &
     \colhead{} &
     \colhead{(TeV)} &
     \colhead{} &
     \colhead{(TeV)}
}
\startdata
\multicolumn{6}{l}{First component} \\
${\rm I}$   & $36300_{-6100}^{+7400}$ & $2.57_{-0.03}^{+0.02}$ &
$468_{-88}^{+95}$ & $1.12_{-0.26}^{+0.33}$ & \nodata \\
${\rm I}'$  & $38000_{-7100}^{+9900}$ & $2.58 \pm 0.03$ &
$447_{-100}^{+103}$ & $1.03_{-0.23}^{+0.33}$ & $4.37_{-2.82}^{+5.87}$ \\
\hline
\multicolumn{6}{l}{Second component} \\
${\rm II}$  & $21500$ & \nodata & $12400$ (fixed)
& \nodata & \nodata \\
${\rm II}'$ & $21500$  & \nodata & $12400$ (fixed) &
\nodata & $5.62_{-2.81}^{+5.60}$ \\
\enddata
\tablenotetext{}{
}
\end{deluxetable*}

\begin{figure}[t]
  \centering
   \includegraphics[width=\linewidth]{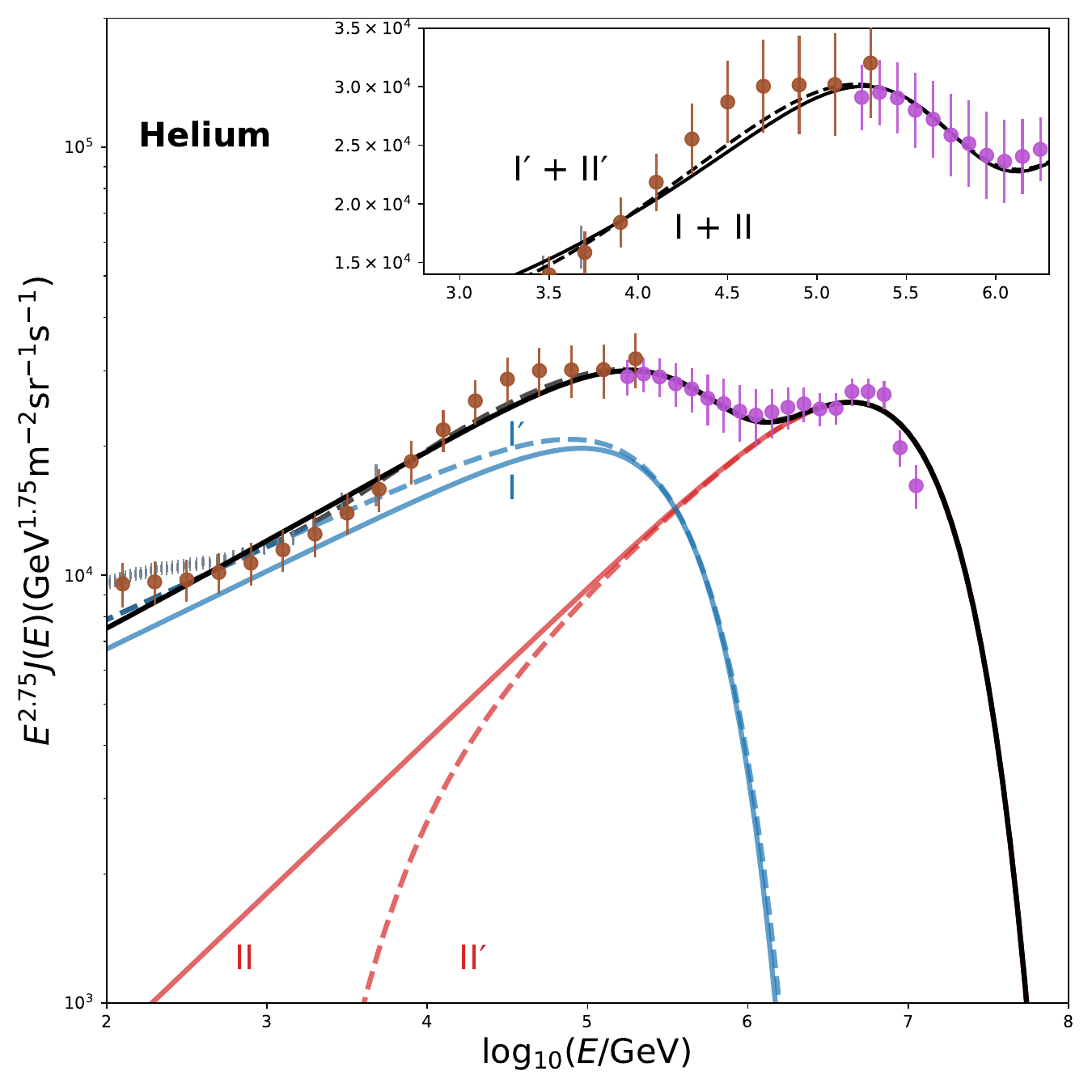}
  \caption{\label{fig5}
CR helium spectrum multiplied by $E^{2.75}$, with fit curves compared to measurements from AMS-02~\citep{AMS:2017seo}, DAMPE~\citep{DAMPE:2025opn}, and LHAASO~\citep{LHAASO:2025mlf}. The blue and red curves represent the contributions of the first (I) and second (II) CR components, respectively. Two alternative assumptions for the extrapolation of the second component toward low energies are shown: (a) a pure power-law extrapolation (curves $\mathrm{I}$ and $\mathrm{II}$) and (b) a scenario with low-energy hardening of the second component (curves $\mathrm{I}'$ and $\mathrm{II}'$), as discussed in the text. The solid and dashed black curves show the resulting total helium spectrum ($\mathrm{I}+\mathrm{II}$) for the (a) and (b) scenarios, respectively. The inset highlights the spectral structure in the multi-TeV energy range.
  }
\end{figure}


\begin{figure}[t]
  \centering
   \includegraphics[width=\linewidth]{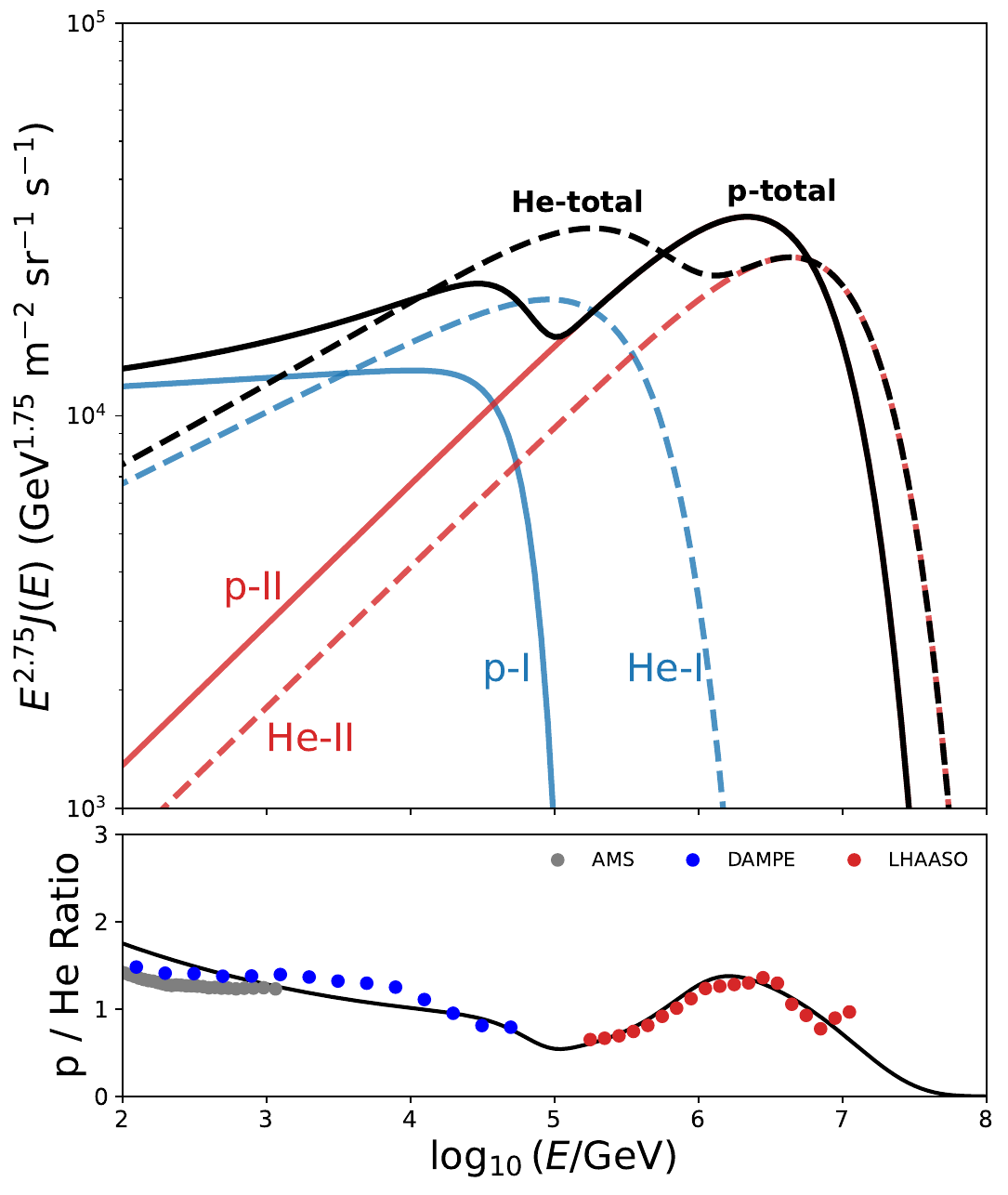}
  \caption{\label{fig6}
  Comparison of the two-component model for cosmic-ray protons and helium. \textbf{Top panel:} Energy spectra multiplied by $E^{2.75}$ for protons (solid curves) and helium (dashed curves). For each species, the blue and red lines show the contributions of the first (${\rm I}$) and second (${\rm II}$) components, respectively. The black lines indicate the total modeled flux, with protons (solid line) and helium (dashed line). \textbf{Bottom panel:} The resulting proton-to-helium (p/He) flux ratio predicted by the model, compared with experimental measurements from the AMS, DAMPE, and LHAASO~\citep{LHAASO:2025mlf}. 
  }
\end{figure}

\subsection{CR  helium spectrum}

The helium measurements from different instruments are compiled in Fig.~\ref{fig:proton_helium_spectrum}(b). As for protons, we apply the two-component approach to the helium data reported by the AMS-02, DAMPE, and LHAASO collaborations. The results are shown in Fig.~\ref{fig5}, and the corresponding best-fit parameters are listed in Table~\ref{tab:spectral_fits_he}.

The fitting function is identical to that used for the proton spectra (see Table~\ref{tab:spectral_fits_he}). As in the proton case, the first CR helium component is obtained from a joint fit to the DAMPE and LHAASO data. For the second component, we do not attempt a dedicated fit to the LHAASO helium data at PeV energies; instead, we scale it relative to the corresponding proton component derived above. Specifically, the second helium component is constructed assuming a rigidity-dependent cutoff energy: its spectral index is fixed to the proton value, and only the normalization is adjusted to match the helium data. The resulting agreement is satisfactory, suggesting that the second components of CR protons and helium may share a common origin.

In contrast, the same does not hold for the first CR helium component. As shown in Table~\ref{tab:spectral_fits_he}, the helium spectrum is harder than the proton spectrum, with a power-law index $\Gamma_{1,\mathrm{He}}$ smaller by $\approx 0.1$ than the corresponding proton index $\Gamma_{1,\mathrm{p}}$ (see Tables~\ref{tab:spectral_fits_1} and \ref{tab:spectral_fits_he}). This behavior is consistent across both variants of the second-component extrapolation.

An even more pronounced difference between the first proton and helium components arises in the cutoff region. The cutoff energy of the helium component, $E_{0,\mathrm{He}}$, exceeds that of the proton component, $E_{0,\mathrm{p}}$, by a factor of $\sim 5$ (see Tables~\ref{tab:spectral_fits_1} and \ref{tab:spectral_fits_he}), significantly larger than the factor of $Z = 2$ expected from simple rigidity scaling. Moreover, the cutoff shapes differ between the two species. The first proton component exhibits a sharp cutoff, characterized by an exponential index $\beta_{1,\mathrm{p}} \sim 2$, whereas the helium component is consistent with a simple exponential cutoff, $\beta_{1,\mathrm{He}} \sim 1$.

The spectral differences between the first proton and helium components may reflect variations in the underlying acceleration mechanisms, potentially favoring heavier nuclei. However, a more natural interpretation is that these differences arise because the two components are produced in different environments. This does not necessarily imply distinct classes of particle accelerators. For example, in core-collapse supernova remnants (SNRs), helium and heavier elements may be accelerated in the reverse shock propagating through the progenitor ejecta, while protons are accelerated in the forward shock  \citep[e.g.,][]{Ptuskin2013}.

As with protons, the two-component model applied to CR helium naturally explains the gradual spectral hardening as the result of the increasing contribution of the second component at lower energies, without invoking additional acceleration- or propagation-related effects. Moreover, the differences between the proton and helium spectra provide a natural explanation for the behavior of the p/He ratio. 

Fig.~\ref{fig6} shows a complex energy dependence of the p/He ratio, spanning more than five decades in energy. Our model reproduces the main features of this behavior, including the decrease and subsequent recovery in the multi-TeV region, which arise from the transition between the two components and their different relative contributions to protons and helium. The maximum in the p/He ratio at PeV energies can be understood as a consequence of the interplay between the two components and their different cutoff energies for protons and helium, and is naturally explained within our framework by the earlier cutoff of the proton spectrum.

Finally, small deviations at the level of a few percent are visible at energies of a few hundred GeV. Although these could be reduced by introducing additional parameters or refinements, we deliberately avoid doing so to preserve the minimal and physically transparent nature of the two-component framework.

{\subsection{Low-energy behavior below 100 GeV}
\begin{figure}[t]
  \centering
   \includegraphics[width=\linewidth]{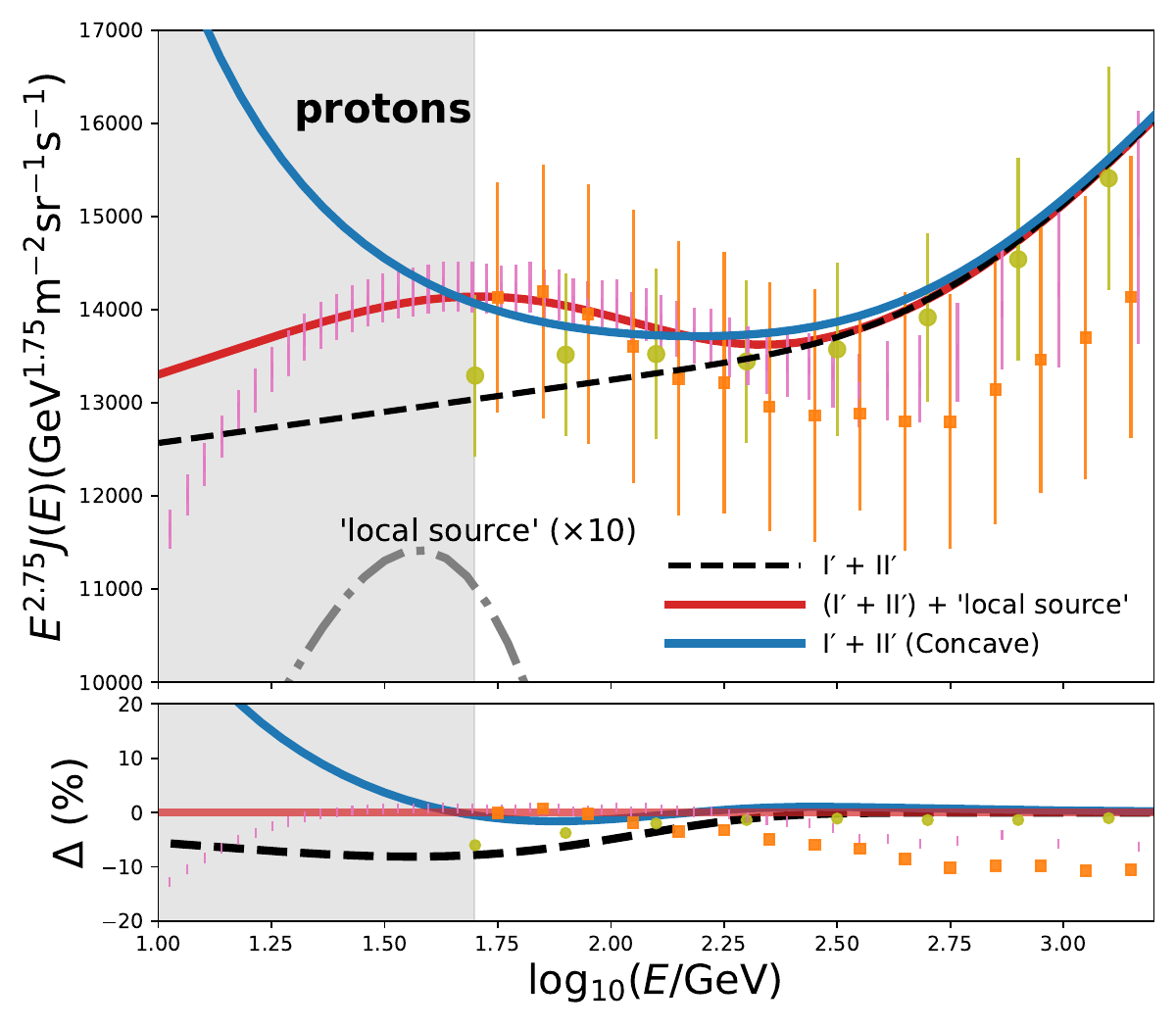}
  \caption{\label{below100} 
Local CR proton spectrum multiplied by $E^{2.75}$
 in the energy range from 10~GeV to 1.5~TeV. Spectral measurements by AMS-02~\citep{Aguilar2015Proton}, DAMPE~\citep{DAMPE:2025opn}, and CALET~\citep{CALET_2022_proton} are shown. The figure illustrates two possible explanations for the mild concavity of the proton spectrum between 50 and 200~GeV, as discussed in the text.
In scenario (a), a contribution from a hypothetical local source is added to the two-component CR model. The dot-dashed grey curve shows the local-source contribution (multiplied by a factor of 10 for clarity), while the solid red curve shows the corresponding total spectrum.
In scenario (b), the first CR component is assumed to possess an intrinsic concave shape, as expected in nonlinear diffusive shock acceleration models. The dot-dashed purple curve represents the resulting spectrum.
The dashed black curve shows the prediction of the original two-component model without any additional low-energy modification.
The grey shaded region below 50~GeV corresponds to the energy range strongly affected by solar modulation.
}
\end{figure}

So far, we have focused on the energy range above 100~GeV. The main reason is that at lower energies, Solar modulation affects the observed CR fluxes, causing significant deviations from the interstellar spectra. Nevertheless, Solar modulation is generally believed to become relatively unimportant above $\sim$50~GeV, making it worthwhile to examine the validity of the proposed framework down to this energy.

From the perspective of the present model, this relatively narrow energy interval should not play a major role. Yet,  in this specific case,  below 200~GeV,  the proton spectrum shows a tendency toward additional steepening, which 
should be addressed. While our proposed model with a pure power-law first component 
agrees with the DAMPE data, it passes slightly below the AMS-02 and CALET spectral points (see   Fig.~\ref{below100}). If the AMS-02 and CALET measurements indeed provide a more accurate description of the proton spectrum below approximately 200~GeV than the DAMPE data—a possibility that cannot presently be excluded—the observed deviations may indicate the presence of additional spectral structure not included in the minimal two-component formulation.

A gradual low-energy steepening could arise from an additional nearby source
contribution whose characteristic energy depends on the source distance, age, injection history, and local diffusion properties. In this sense, as with the multi-TeV spectral feature, 
the low-energy structure
may also reflect the contribution of local source populations, albeit of a different physical origin.

Alternatively, such behavior may reflect the expected concavity of spectra produced within
nonlinear diffusive shock acceleration models. In this context, it is worth noting that similar
scenarios were discussed by  \cite{Ptuskin2013} in connection with the PAMELA data, which are themselves consistent with AMS-02 measurements. 

Both options are demonstrated qualitatively in  Fig.~\ref{below100}.  Interestingly, the corresponding spectral structure appears to be less pronounced in the helium spectrum. 
Within the two scenarios discussed above, this could be explained either by a high p/He ratio in the local source population or by different source environments contributing to the proton and helium populations of the first CR component.

\section{Astrophysical Implications}

In this section, we discuss the astrophysical implications of the proposed minimal two-component cosmic-ray framework, focusing on the nature of the source populations that may contribute to the two distinct proton components identified in the spectral analysis. The phenomenological decomposition, by itself, does not allow robust conclusions about individual sources, given the superposition of potential contributions from many objects distributed throughout the Galaxy. Nevertheless, the inferred spectral properties provide meaningful constraints for assessing the plausibility of different classes of Galactic accelerators.

\subsection{SNR origin of the first CR component}

The coherent description of the Galactic CR spectrum from GeV to PeV energies within the two-component framework does not require a fundamental revision of the standard SNR paradigm for the origin of Galactic CRs  \cite[for a recent review, see][]{Bykov2025}. Rather, it supports the validity of this paradigm up to $\sim 100~\mathrm{TeV}$, beyond which the second (“PeVatron”) component becomes dominant. In this context, the two-component picture naturally alleviates long-standing tensions associated with the theoretical difficulty of accelerating protons beyond $\sim 100~\mathrm{TeV}$ in young SNRs within the standard DSA framework \citep{LagageCesarsky1983a,Hillas2005}. The presence of the second component can, in principle, still be accommodated within the SNR paradigm, for example, if PeV particles are produced in a special class of SNRs. However, such sources may compete with other potential PeVatrons, notably microquasars.

The two-component framework, as far as the first component is concerned, preserves the well-established aspect of this paradigm—the production of relatively hard source spectra and their subsequent steepening due to energy-dependent propagation in Galactic magnetic fields.
This interpretation also helps to reconcile cosmic-ray data with the somewhat unexpectedly steep multi-TeV $\gamma$-ray spectra observed above $\sim 1~\mathrm{TeV}$ in many young SNRs \citep[e.g.,][]{AharonianYangWilhelmi2019}. These include both classical remnants, such as SN~1006 and Cas~A, and more peculiar objects, such as RX~J1713.7$-$3946. However, caution is required when interpreting these $\gamma$-ray observations as direct probes of proton acceleration. Inverse Compton (IC) scattering by relativistic electrons is an efficient radiation mechanism at TeV energies and may dominate over hadronic $\gamma$-ray production, particularly in environments where the magnetic field strength does not significantly exceed $\sim 10~\mu\mathrm{G}$. As a result, current $\gamma$-ray measurements alone do not allow firm conclusions regarding the maximum energies attained by accelerated protons.

Moreover, even in the case of a hadronic origin of $\gamma$-ray emission, the observed spectral steepening at TeV energies is likely influenced by the escape of the highest-energy particles from the acceleration region, including in relatively young remnants. In this context, $\gamma$-ray emission from regions surrounding SNRs may provide a more direct and less biased probe of the maximum proton energies \citep{Aharonian2013SNRreview}. In the absence of comprehensive morphological studies of young SNRs and their environments, current $\gamma$-ray observations alone cannot be considered reliable tracers of the maximum proton energies.

The strongest argument in favor of SNRs as the dominant source population responsible for Galactic CRs remains, as for decades, the energetic requirement. The total power available in the form of kinetic energy carried by SNR shocks can reach $\sim 10^{42},\mathrm{erg,s^{-1}}$, and it has long been assumed that $\sim 10 \%$ of this energy is converted into CRs \citep{GinzburgSyrovatskii1964}. This phenomenological estimate is broadly supported by non-linear shock acceleration theory \citep[see][]{MalkovDrury2001}, which generically predicts efficient conversion of kinetic energy into CRs, thereby reinforcing the role of SNRs as a major contributor to Galactic CRs.

The relatively low cutoff energy, $E_{0,1} \lesssim 100~\mathrm{TeV}$, inferred for the first CR proton component is consistent with conservative expectations for the limited potential of DSA operating in the bulk of “standard” SNRs. The super-exponential cutoff required by the data, with $\beta_1 \sim 2$, can be interpreted as evidence for a rapid onset of particle escape near the maximum attainable energy. Such behavior may arise from a sudden loss of particle confinement, in which the diffusion length of the highest-energy particles becomes comparable to the spatial extent of the acceleration region, thereby enabling efficient upstream escape. One plausible physical mechanism is the breakdown of self-generated magnetic turbulence: above a critical energy, the accelerated particles may no longer sustain the turbulence required for effective confinement, leading to a sharp suppression of the confined proton spectrum. In this regime, insufficient turbulence not only limits the maximum attainable energy to values below $\sim 100~\mathrm{TeV}$, but also naturally produces a very sharp spectral cutoff. Without entering into the details of this highly complex and model-dependent process, we emphasize that, at a phenomenological level, the data require the first CR component to terminate abruptly, favoring a super-exponential cutoff with $\beta_1 \gtrsim 2$.

Although we consider CR protons and helium within the same two-component framework, it can, in principle, be extended to heavier nuclei. However, the spectra of different species need not be similar, as they may originate in different source environments with distinct compositions and acceleration conditions.

The spectra derived for CR helium reveal a notable result. The helium population of the second component agrees well with the proton spectrum when expressed in terms of rigidity, indicating that these components likely share a common origin and dominate the CR flux at PeV energies. In contrast, the situation for the first helium component is markedly different. The spectrum of the first helium component is not only significantly harder than its proton counterpart ($\Delta \Gamma \approx 0.1$), but also extends to higher energies ($E_{0,\mathrm{He}} \approx 5,E_{0,\mathrm{p}}$), exceeding the expectation from simple rigidity scaling. Moreover, the cutoff shape differs substantially: the helium component exhibits a smoother cutoff, whereas the proton component is characterized by a super-exponential suppression.

This behavior is most naturally interpreted as evidence that the p-I and He-I components originate in different source environments with distinct compositions and acceleration conditions. For example, this may suggest that the He-I component is produced in metal-rich SNRs or in the reverse shocks of core-collapse SNRs propagating through the ejecta of massive-star progenitors.

Another distinct source population -- young stellar clusters (YSCs) and, more broadly, star-forming regions -- has also long been discussed as a potential contributor to Galactic CRs. Particle acceleration in colliding stellar winds and in the multiple shocks driven by supernova explosions within such systems was proposed decades ago as an alternative to isolated SNRs \citep{CesarskyMontmerle1983,Bykov_superbubbles}. Recent reports of extended diffuse $\gamma$-ray emission associated with several stellar clusters, with hard spectra extending up to $\sim 10~\mathrm{TeV}$, have renewed interest in high-energy processes operating in YSCs \citep{AharonianYangWilhelmi2019}. This is not unexpected, given that the total mechanical power injected by stellar winds and supernovae in massive clusters is only a factor of a few below the kinetic power released by Galactic supernova explosions, rendering YSCs energetically viable contributors to the Galactic CR budget.

In the proposed two-component framework, YSCs and superbubbles are treated as complementary acceleration environments rather than primary sources of the first CR population. While their integrated mechanical power may substantially contribute to CR production at lower energies, their key role is to sustain particle acceleration under extreme, long-lived conditions. This naturally links these systems to the second CR component, which dominates above $\sim 100~\mathrm{TeV}$ and requires acceleration sites that exceed the capabilities of isolated SNRs evolving in typical interstellar environments.

\subsection{PeVatrons and the second CR component}

The detection of dozens of ultra-high-energy (UHE; $E \gtrsim 100~\mathrm{TeV}$) $\gamma$-ray sources associated with a variety of astronomical objects \citep{LHAASO-catalog} indicates that the Galaxy hosts a diverse population of PeVatrons rather than a single dominant source class. These objects are natural candidates for contributing to the second CR proton component.
In the following, we briefly discuss three classes of astrophysical sources relevant to this component. We begin by revisiting the role of supernova remnants, focusing on the possibility that a subset of remnants, under favorable conditions, can operate as PeVatrons. We then consider young stellar clusters and extended star-forming regions as environments where collective effects and clustered supernova activity may enhance particle acceleration to PeV energies. Finally, we discuss microquasars, which were traditionally underestimated as cosmic-ray accelerators but have recently attracted renewed attention as viable Galactic PeVatron candidates in light of recent UHE $\gamma$-ray discoveries.

\subsubsection{Old SNRs as smoking UHE $\gamma$-ray guns}

While supernova remnants are proposed here as the dominant contributors to the first CR proton component, characterized by a sharp decline above $\sim 100~\mathrm{TeV}$, it remains possible that a subset of SNRs operates as proton PeVatrons and contributes non-negligibly to the second CR component. Under favorable conditions, protons may be accelerated to PeV energies during the very early stages of SNR evolution, when shock velocities can reach a significant fraction of the speed of light ($v_{\rm sh} \gtrsim 0.03,c$) and the magnetic field is strongly amplified, for example through CR-induced instabilities. In such environments, the combination of strong shocks, enhanced magnetic turbulence, and high ambient densities can raise the maximum attainable particle energy into the PeV range \citep[e.g.][]{Ptuskin2012,Bell2013,Blasi2013Review,Bykov2025}.

The short duration of the PeVatron phase and the rapid escape of the highest-energy particles from the remnant make it inherently difficult to identify such objects through direct $\gamma$-ray observations of young SNR shells. The absence of UHE $\gamma$-ray detections from young SNRs should therefore not be regarded as a decisive argument against their ability to act as PeVatrons.

A more promising approach is to conduct indirect searches for so-called “smoking guns,” namely the detection of $\gamma$-ray emission from cosmic-ray protons and nuclei that have escaped their accelerator and subsequently interact with nearby giant molecular clouds (GMCs). In this scenario \citep{AhAt1996clouds}, $\gamma$-ray emission may be observed thousands of years after the accelerator itself has faded away. In the context of supernova remnants, this scenario has long been recognized as a viable method for identifying SNRs operating as proton PeVatrons \citep{GaAh-SNR-GMC}. Before fully diffusing and mixing into the Galactic cosmic-ray “sea,” these runaway particles can produce $\gamma$-ray emission in nearby massive gas complexes, with spectra that differ significantly from those produced in the SNR shell and may extend well beyond $100~\mathrm{TeV}$. The realization of this scenario can significantly increase the number of SNR-related systems detectable at ultra-high energies. 

As suggested in the LHAASO Collaboration discovery paper \citep{LHAASO-pevatrons}, some of the reported UHE $\gamma$-ray sources may indeed be associated with SNR–GMC complexes. Subsequent dedicated studies aimed at identifying UHE $\gamma$-ray emitters from the first LHAASO source catalog \citep{LHAASO-catalog} have revealed several candidate associations with GMCs located near middle-aged SNRs (see, e.g., \citealt{AlisonSilvia}). The spectra of these sources extend beyond $100~\mathrm{TeV}$, implying that the energies of the parent particles, most likely protons, must reach the PeV scale. While these associations are highly suggestive, they do not yet constitute unambiguous proof of a causal link between UHE $\gamma$-ray sources and middle-aged SNRs. Future UHE $\gamma$-ray observations with substantially improved angular resolution, combined with comprehensive multiwavelength studies, will be essential for establishing whether SNRs can indeed operate as proton PeVatrons.

\subsubsection{Star-forming regions as sites of proton PeVatrons}

Stellar clusters and, more broadly, star-forming regions (SFRs) were identified long ago as efficient CR production sites \citep[e.g.,][]{CesarskyMontmerle1983}. These regions are characterized by powerful stellar winds from massive stars, which generate enhanced magnetic fields and drive strong turbulence. When combined with shocks produced by ongoing supernova activity, these conditions may establish environments more conducive to proton acceleration to PeV energies than those 
found in isolated SNRs.

Recent $\gamma$-ray observations have revealed diffuse UHE $\gamma$-ray structures associated with several prominent stellar complexes in the Milky Way, including the powerful young stellar cluster Westerlund~1 \citep{HESS_West1}, the Cygnus Cocoon \citep{LHAASO_CygnusBubble}, and the active star-forming region W43 \citep{LHAASO_W43}. Of particular interest is the Cygnus Cocoon, whose $\gamma$-ray spectrum extends into the PeV range \citep{LHAASO_CygnusBubble}, implying the presence of multi-PeV protons. A leptonic interpretation of the UHE $\gamma$-ray emission in terms of inverse Compton (IC) scattering is strongly disfavored, given the rapid synchrotron losses that prevent electrons from reaching PeV energies in typical magnetic-field environments.

Particle acceleration within stellar clusters may proceed via two main scenarios. The first involves acceleration at shocks generated by colliding stellar winds, which can efficiently energize particles up to multi-TeV energies. However, detailed theoretical studies indicate that reaching the PeV domain through stellar-wind interactions alone is highly challenging \citep[e.g.,][]{Morlino2019,Vieuetal2022,Harer2025}.
Alternatively, supernovae occurring within stellar clusters may play a dominant role. In this scenario, supernova shocks propagate into a pre-existing, turbulent and strongly magnetized medium created by stellar winds, which can significantly enhance acceleration efficiency and relax constraints on the maximum attainable particle energy \citep{Vieuetal2022}. This mechanism is particularly relevant for the Cygnus region, notably the stellar association Cygnus~OB2. In this picture, the stellar association itself does not act as the PeV proton accelerator; rather, it provides favorable conditions that enable a powerful supernova -- possibly one that occurred a few $\times 10^{4}$ years ago -- to accelerate protons up to PeV energies, potentially accounting for the UHE $\gamma$-ray emission observed toward the Cygnus region \citep{Harer2025}.

Nevertheless, even under favorable conditions, explaining the extension of the Cygnus Cocoon spectrum to PeV energies remains challenging, suggesting that additional or alternative PeVatron sources may contribute to the observed PeV $\gamma$-ray emission.

\subsubsection{Microquasars as potential PeVatrons}

Microquasars are luminous X-ray binaries hosting a stellar-mass black hole or neutron star, which launch collimated, accretion-powered outflows. The kinetic power of these jets can episodically reach $L_{\rm K} \simeq 10^{38}$--$10^{40}~\mathrm{erg,s^{-1}}$, comparable to or even exceeding the Eddington luminosity of the compact object. Such extreme power output provides favorable conditions for transient PeVatron activity \citep{Enrico_mQSO,Jishuang_mQSO,Kaci_mQSO, Zhang:2025tew}.

Within the framework of ideal MHD, the maximum energy attainable by a proton is constrained by the electric potential drop across the system and can be estimated as \citep{Jishuang_mQSO}
\begin{equation}
E_{\rm max} \approx 10
\left(\frac{L_{\rm K}}{10^{38}~\mathrm{erg,s^{-1}}}\right)^{1/2}
\left(\frac{\sigma v}{c}\right)^{1/2}~\mathrm{PeV},
\label{Emax}
\end{equation}
where $v$ is the jet velocity, $c$ is the speed of light, and $\sigma$ is the magnetization parameter characterizing the fraction of kinetic energy converted into Poynting flux. Thus, microquasars in high-accretion states, with relativistic jets of mechanical power $L_{\rm K} \gtrsim 10^{39}~\mathrm{erg,s^{-1}}$ and moderate magnetization ($\sigma \gtrsim 0.1$), can in principle accelerate protons to energies exceeding $\sim 10~\mathrm{PeV}$.

Recent detections of diffuse UHE $\gamma$-ray emission associated with several powerful Galactic microquasars \citep{HAWC_SS433,HESS_SS433,HAWK_V4641Sgr,HESS_V4641Sgr,LHAASO_mQSOs} place these systems firmly among the growing class of Galactic PeVatron candidates. In the case of the most powerful system in the Galaxy, SS~433, the multi-TeV emission exhibits a pronounced energy-dependent morphology \citep{HESS_SS433}, which is best explained by IC scattering of relativistic electrons. explained by IC scattering of relativistic electrons.
However, the detection by LHAASO of UHE $\gamma$ rays extending beyond $100~\mathrm{TeV}$ \citep{LHAASO_mQSOs} indicates the presence of an additional high-energy emission component that is difficult to reconcile with a purely leptonic origin and instead favors hadronic processes. Even more compelling is the detection of $\gamma$ rays with energies well above $100~\mathrm{TeV}$ from V4641~Sgr \citep{LHAASO_mQSOs}, which is challenging to explain within standard inverse Compton scenarios.

Meanwhile, the detection by LHAASO of UHE $\gamma$ rays extending beyond
$100~\mathrm{TeV}$ \citep{LHAASO_mQSOs} indicates the presence of an additional
high-energy emission component that is difficult to reconcile with a purely
leptonic origin and more naturally points to hadronic processes. Even more
compelling is the detection of $\gamma$ rays with energies well above
$100~\mathrm{TeV}$ from V4641~Sgr \citep{LHAASO_mQSOs}, which is difficult
to accommodate within any reasonable inverse Compton framework.

Finally, the most compelling evidence for proton acceleration to PeV energies is provided by the discovery of PeV $\gamma$ rays from the microquasar Cygnus~X--3 by the LHAASO Collaboration \citep{LHAASO_CygX3}. The observed flux variability and its modulation with the 4.8~h orbital period demonstrate that the parent particles are accelerated within the extremely compact binary system itself, leaving little room for a leptonic interpretation of the $\gamma$-ray emission. The modulated signal, characterized by an unusually hard spectrum extending up to $\sim 3~\mathrm{PeV}$, is best explained by photomeson production resulting from interactions of protons with energies exceeding $\sim 10~\mathrm{PeV}$ with the intense optical/UV radiation field of the massive companion star \citep{LHAASO_CygX3}. This discovery can therefore be regarded as the first unambiguous identification of a Galactic proton PeVatron.

From a theoretical standpoint, microquasars offer two plausible sites for proton acceleration to extreme energies. The first involves internal shocks within the compact, relativistic jets formed inside the binary system, as inferred for Cygnus~X--3. The second corresponds to jet-termination shocks developing at distances of tens of parsecs from the source, as suggested for V4641~Sgr and possibly also for Cygnus~X--3, if the extended UHE $\gamma$-ray emission observed toward the Cygnus region originates from a halo powered by the microquasar outflow\footnote{This interpretation, as an alternative to a link with the Cygnus~OB2 stellar association, cannot be excluded \emph{a priori}. Moreover, it alleviates the theoretical tension associated with accelerating multi-PeV protons in this powerful but loosely bound stellar association.}. The substantial kinetic power transported by relativistic jets renders microquasars natural candidates for both persistent and transient PeVatron activity.

The energy required to sustain the observed PeV CR proton flux, $\dot{W}_{\rm CR}(\geq 1~\mathrm{PeV})$, is estimated to be $\simeq 10^{38}$–$10^{39}\,\mathrm{erg\,s^{-1}}$, depending on the highly uncertain confinement time of PeV protons inside the Galaxy. This implies that even a single or a small number of powerful microquasars could, in principle, supply the PeV CR flux, given that proton acceleration to PeV energies requires a comparable power per source. Indeed, it has been shown that the presence of only $\sim 10$ active powerful microquasars in the Galaxy at any given time is sufficient to reproduce CR and $\gamma$-ray data within a self-consistent framework \citep{Kaci_mQSO}; see also \cite{Zhang:2026igt} for a scenario in which the contribution is dominated by a few nearby microquasar remnants. At the same time, the total power released by the microquasar population remains well below the injection rate required to sustain the overall Galactic CR luminosity, $\dot{W}_{\rm CR,tot} \simeq 10^{41}\,\mathrm{erg \,s^{-1}}$. 

Thus, within our two-component cosmic-ray proton framework, microquasars are not expected to provide a significant contribution to the low-energy component. At the same time, the extension of the CR proton spectrum beyond $10~\mathrm{PeV}$ (see Fig.~\ref{fig2}) raises the question of whether microquasars could also be responsible for this component of the spectrum, at least up to $100~\mathrm{PeV}$. 
It follows from Eq.~(\ref{Emax}), achieving such energies would require jet powers that appear unrealistically large, exceeding $\sim 10^{40}\,\mathrm{erg\,s^{-1}}$.

If the CR proton component extending well beyond $10~\mathrm{PeV}$ is of Galactic origin, individual sources or phenomena more powerful than microquasars and supernova explosions are required. In this context, the supermassive black hole at the Galactic Center, Sgr~A$^\ast$, appears to be the only viable candidate within the Milky Way. Although Sgr~A$^\ast$ currently accretes at a very low rate, corresponding to $\sim 10^{-8} L_{\rm Edd}$ for a black hole of mass $\sim 4 \times 10^{6}\,M_\odot$, its past activity may have been orders of magnitude higher. During such active phases, Sgr~A$^\ast$ could in principle satisfy the conditions required to accelerate protons to energies approaching, or even exceeding, $100~\mathrm{PeV}$. Interestingly, Sgr~A$^\ast$ has been proposed as a Galactic PeVatron injecting protons into the Central Molecular Zone to explain the diffuse multi-TeV $\gamma$-ray emission observed from that region \citep{GC_Pevatron}. This hypothesis requires confirmation through $\gamma$-ray observations extending into the PeV energy range.

\section{Summary}

Recent high-precision measurements of the CR proton spectrum reveal significant deviations from a simple power-law behavior, including (i) progressive spectral hardening above $\sim 200~\mathrm{GeV}$, (ii) a pronounced excess in the $10$--$30~\mathrm{TeV}$ range followed by a sharp turnover around $\sim 100~\mathrm{TeV}$, and (iii) a broad spectral structure extending from $\sim 0.1~\mathrm{PeV}$ to several PeV. These features cannot be readily explained within a single-component framework based solely on standard acceleration and propagation effects.

In this work, we have shown that the entire proton spectrum from GeV to PeV energies can be consistently described by the superposition of two broad Galactic CR proton populations. In this minimal two-component framework, the observed spectral complexity arises naturally from the transition between a low-energy component with a sharp cutoff at tens of TeV and a second, harder component that dominates above $\sim 100~\mathrm{TeV}$ and extends to PeV energies.

The two-component model can be successfully applied to the CR helium population. It requires a harder ($\Delta \Gamma \approx 0.1$) first helium component with a power-law spectrum than that of the first proton component, extending effectively to several hundred TeV. At the same time, the second helium component scales with the proton spectrum as a function of rigidity. The superposition of these two components provides a consistent description of both the helium spectrum and the energy dependence of the p/He ratio from 100 GeV to 10 PeV.  In the 50--200 GeV energy range, the proton spectrum may exhibit a mild concavity relative to a pure power-law first component. If this feature reflects the underlying interstellar CR spectrum rather than local source contributions or residual systematic effects, it may arise naturally from the characteristic concavity predicted by nonlinear diffusive shock acceleration models.

The proposed model reproduces all major spectral features without invoking highly contrived acceleration or propagation scenarios. Moreover, the present data do not require a dominant contribution from a nearby (local) source to explain the observed proton spectrum. While such sources cannot be excluded in principle, the two-component interpretation provides a population-based explanation at the Galactic scale and is therefore favored on grounds of simplicity. In this sense, the proposed model follows the spirit of {\it Occam's razor}, introducing no additional ingredients beyond those minimally required by the data.

\begin{acknowledgments}
B.T.Z. is supported in China by the National Key R\&D program of China under the grant 2024YFA1611402.
\end{acknowledgments}

\clearpage

\bibliography{references}

\end{document}